\documentclass[aps,prl,reprint,superscriptaddress]{revtex4-1}
\usepackage{amsmath,amsfonts,amssymb,graphicx,hyperref,epstopdf,xcolor,tikz,scalerel,natbib}
\usepackage{mathptmx,textcomp}
\usepackage[T1]{fontenc}
\usepackage[utf8]{inputenc}

\definecolor{lime}{HTML}{A6CE39}
\DeclareRobustCommand{\orcidicon}
{
	\begin{tikzpicture} 
	\draw[lime, fill=lime] (0,0) circle [radius=0.15] node[white] {{\fontfamily{qag}\selectfont \tiny ID}};
	\draw[white, fill=white] (-0.0625,0.095) 	circle [radius=0.007];
	\end{tikzpicture}
	\hspace{-2.2mm}
}
\newcommand\orcidID[1]{\href{https://orcid.org/#1}{\orcidicon}}

\newcommand{\DDT}[1]{\frac{d#1}{dt} }
\newcommand{\DDTa}[1]{\frac{d#1}{d\eta} }
\newcommand{\pddt}[1]{ \partial#1/\partial t}
\newcommand{\pddz}[1]{ \partial#1/\partial z}
\newcommand{\Pdd}[2]{\frac{\partial#1}{\partial #2} }
\newcommand{\be}{\begin {equation}}
\newcommand{\ee}{\end {equation}}
\newcommand{\beqa}{\begin {eqnarray}}
\newcommand{\eeqa}{\end {eqnarray}}
\newcommand{\mb}{\mathbf}
\hypersetup{colorlinks,citecolor=blue,filecolor=black,linkcolor=blue,urlcolor=blue}

\begin{document}

\title{Complete characterization of ultra-intense laser pulses in radiation damping regime}
\author{Amol R. Holkundkar\orcidID{0000-0003-3889-0910}}
\email{amol.holkundkar@pilani.bits-pilani.ac.in}
\affiliation{Department of Physics, Birla Institute of Technology and Science - Pilani, Rajasthan,
333031, India}

\author{Felix Mackenroth\orcidID{0000-0002-2456-5917}}
\affiliation{Max Planck Institute for the Physics of Complex Systems, Dresden, Germany}

\date{\today}

\begin{abstract}
We report the first closed, analytical expression for the scattering angle of an electron bunch ponderomotively scattered from a counter-propagating, ultra-intense laser pulse, also accounting for radiation reaction (RR). The found formulation depends nontrivially on the laser intensity, pulse duration, beam waist, and energy of the electron bunch. For various laser and bunch parameters the proposed formula is in excellent quantitative agreement with full, relativistic test particle simulations in a realistic electromagnetic field configuration of a focused laser pulse. We also demonstrate how in the radiation dominated regime a simple rescaling of our model's input parameters yields excellent quantitative agreement with numerical simulations based on the Landau-Lifshitz model. Finally, we discuss how the model can be applied for an in-situ characterization of current and future ultra-high power laser systems, but also to experimentally probe fundamental properties of RR during ultra-intense laser electron interaction.  

\end{abstract}


\maketitle
 
The technological developments in recent decades by virtue of chirped pulse amplification technique \cite{Strickland1985219}, resulted in an exponential growth in available peak laser powers and intensities \cite{PhysRevSTAB.5.031301,Gales_etal_2018}. With these breakthroughs, laser fields of intensities as high as $\gtrsim 10^{23}$ W/cm$^2$ seem to be feasible in the near future with state of the art laboratories such as the Vulcan 20 PW upgrade \cite{Vulcan}, the Extreme Light Infrastructure (ELI) Facility \cite{ELI}, the XCELS project \cite{XCELS}, and Apollon project \cite{10.1017/hpl.2014.41}. In the parameter regime explored by these high power lasers it is crucial to study the broadening of the emission spectra \cite{PhysRevAccelBeams.19.114701}, electron-positron pair production \cite{PhysRevD.93.045010}, strong field QED effects \cite{doi:10.1063/1.5086933} and many other contemporary problems. For a thorough investigation of these effects, finely controlled experiments are essential, implying the necessity for an accurate characterization of the employed high-intensity laser pulses. Conventional solid-state devices, however, are not adequate to measure, e.g., peak intensity, pulse duration and beam waist of an ultra-intense laser pulse directly in its focus, as the strong electric field disintegrates the optical components. The standard optical metrology is thus merely capable of running at reduced power, which does not necessarily map the field structure at full intensities correctly \cite{10.1038/s41598-019-55949-3}.

There were various techniques proposed to directly characterize ultra-intense laser pulses. The pulse duration of an ultra-intense pulse can be measured directly in its focus by detecting the radiation pattern emitted by a chirped electron bunch \cite{10.1038/s41598-019-55949-3}.  Lately the measurement of the beam waist of an ultra-intense laser pulses is also carried out by  pondermotive scattering of an electron bunch \cite{Felix2019_NJP}. Nonlinear Thomson/Compton scattering and associated attosecond bursts of electromagnetic radiation are also reported as potential diagnostic tool for the characterization of the ultra-intense laser pulses \cite{10.1103/physrevaccelbeams.21.114001,PhysRevLett.105.063903}. The carrier envelope phase (CEP) of an ultra-intense laser has also been  characterized  recently \cite{PhysRevLett.120.124803}. Furthermore,  atomic ionization was also reported to hold potential for a spatio-temporal characterization of intense laser pulses \cite{10.1088/1361-6587/ab1d3e,PhysRevA.99.043405}. Generally an experiment is designed to explore a physical parameter be it the laser amplitude, it's duration, beam waist or a carrier envelope phase. The analytical treatment of the relativistic pondermotive force is also reported in Ref. \cite{PhysRevE.58.3719}, but its potential as a diagnostic tool for complete characterization of ultra-intense laser is not yet explored. 

In this Letter, we derive an analytical expression for the pondermotive scattering angle of a laser-scattered, counter-propagating electron bunch. The proposed formulation is found to accurately predict the scattering angles as a function of laser peak intensity, its duration and the beam waist. The generic nature of the proposed expression is found to be valid even for the cases when the radiation reaction can not be ignored (i.e. the energy lost by the charge particle in the form of radiation). The scattering angles will depend on the final energy of the electrons after interacted with the ultra-intense laser, rather than the discrete interaction dynamics inside the pulse. This study would not only be crucial for experimentalists, optimising the design and developing next generation ultra-intense lasers, but would also be important for the verification and understanding of radiation reaction (RR) during the interaction of ultra-intense lasers with relativistic electrons \cite{PhysRevX.8.011020,PhysRevX.8.031004}. 

We consider the interaction of a relativistic electron with a counter-propagating focused Gaussian laser pulse. The laser is considered to be polarized along the $x$ direction and propagates along the $z$ direction. Time and space coordinates are  normalized with respect to laser frequency ($\omega$) and the wave vector ($k$) respectively ($kx'\rightarrow x$ and $\omega t' \rightarrow t$, where $x
'$ and $t'$ are space and time coordinates in SI units). The laser field is expressed in the following vector potential form
\be  \mb{A} = \frac{a_0}{\sqrt{1 + z^2/z_r^2}}  \exp\left[\frac{-r^2}{w_0^2(1+z^2/z_r^2)}\right] 
g(\eta) \cos(\eta)\ \mb{e_x}  \label{vecPot} \ee 
where, $a_0 = |e|A_0/m_ec$ is the dimensionless amplitude of the vector potential, $e$ ($m_e$) are charge (mass) of the electron, $c$ is speed of light, $\eta = t - z + \phi_0$, $\phi_0$ is  phase constant considered to be zero, $w_0 \equiv k w_0'$ (where $w_0'$ is the beam waist in SI unit) is the \textit{dimensionless} beam waist, $z_r \equiv w_0^2/2$ is the Rayleigh range, $r \equiv \sqrt{x^2+y^2}$ is the radial distance from the beam axis, and $g(\eta) = \exp[-4\ln(2) \eta^2 \tau_0^{-2}]$ is the Gaussian envelope with the FWHM pulse duration $\tau_0$. It should be noted that Eq. \ref{vecPot} closely resembles realistic focused electromagnetic fields \cite{PhysRevSTAB.5.101301}. In the latter part of the manuscript, we intend to compare the results of our analysis with full, relativistic test particle simulations adhering to actual focused electromagnetic fields beyond the paraxial approximation \cite{PhysRevSTAB.5.101301}. The electron bunch is considered to be a relativistic electron sheet (RES) \cite{10.1103/physrevstab.14.070702,PhysRevLett.104.234801}. The use of RES in the context of nonlinear Thomson and Compton scattering has been studied quite extensively in the past. These RES can be obtained by shining a high-intensity laser on to a thin metallic foil ($\sim 1$ nm), electrons from the foil are dragged with the laser through relativistic self induced transparency. These electrons also contain a transverse momentum component because of the polarization of the laser pulse. This transverse momentum can be mitigated by introducing a secondary thick layer, which tends to completely reflect the laser pulse, but allows  RES to emerge on the other side without any transverse momentum \cite{PhysRevLett.104.234801}.  

The relativistic equations of motion of an electron under the influence of the laser fields is given by ($e = m_e = c = 1$),
$d\mb{p}/dt = -\mb{E} - \mb{v}\times\mb{B}$, where $\mb{p} = \gamma\mb{v}$ is the relativistic momentum, $\mb{v}$ and $\gamma$ are respectively the velocity and associated relativistic factor of the electron, $\mb{E} = -\pddt{\mb{A}}$ and $\mb{B} = \nabla \times \mb{A}$ are the electromagnetic fields of the laser. The force equation along the polarization direction can be written by noting  $A_y = A_z = 0$ as
\be 
\DDT{\tilde{p}_x} = - \upsilon_x \Pdd{A_x}{x} = \frac{p_x}{\gamma} \frac{2 x}{w_0^2(1 + z^2/z_r^2)} A_x, \label{pxtilde}
\ee
where, $\tilde{p}_x \equiv p_x - A_x$ and $\upsilon_x = p_x/\gamma$ is the electron's velocity along the $x$ direction. It should be noted that there will no external force on the electron once the laser is over or it leaves the focus, and hence in the first order approximation we can consider $p_x \sim A_x$, however $\tilde{p}_x$ is just a correction term which arises because of the gradient of the vector potential or fields for that matter. The displacement along the polarization direction can easily be estimated to be $x \sim x_0 + p_x \eta/\gamma$, with $x_0$ being the initial position of the electron in the bunch.  Using these simple approximations Eq. \ref{pxtilde} simplifies to,
\be \DDTa{\tilde{p}_x} \sim \frac{2}{\gamma \DDT{\eta}} \frac{A_x^2}{w_0^2(1 + z^2/z_r^2)} 
\Big[x_0 + \frac{A_x \eta}{\gamma}\Big] \label{ptildeeta}.\ee
\begin{figure}[b]
\includegraphics[totalheight=0.9\columnwidth]{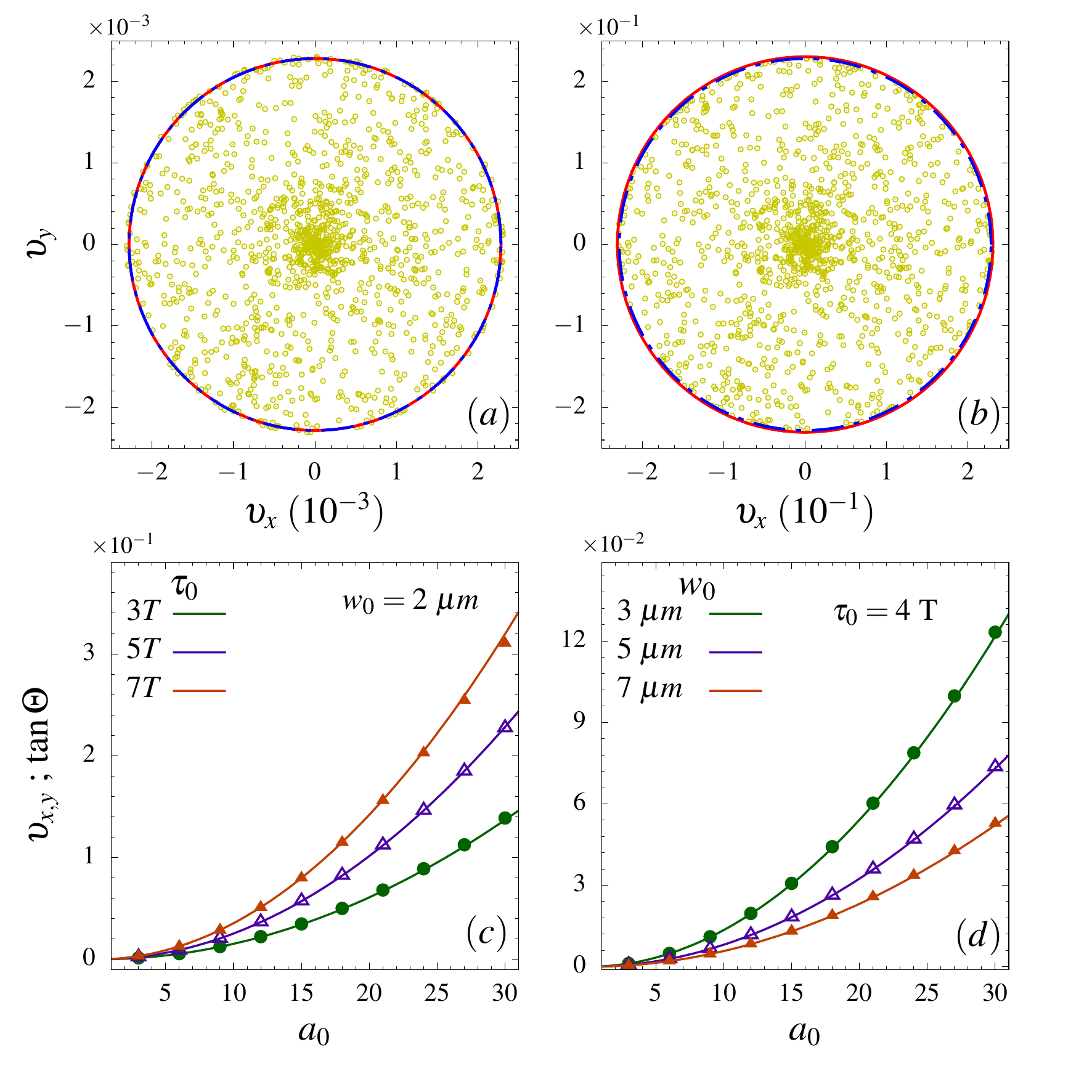}
\caption{The analytical estimates of scattering angles by Eq. \ref{thnorr} are compared with the full relativistic test particle simulations. The final transverse velocity of the electron bunch after the interaction is presented for the peak laser amplitude $a_0 = 3$ (a) and $a_0 = 30$ (b), the laser is considered to be having a FWHM pulse duration of 5 cycles and beam waist of 2 $\mu$m. The dots in (a) and (b) represent the bunch electrons, and the enclosing circle is the theoretically predicted [by Eq. \ref{thnorr}] scattering angle. The numerically calculated scaling of the scattering angles with the peak laser amplitude for a fixed beam waist of 2 $\mu$m and FWHM pulse duration of 3, 5 and 7 cycles are compared with the theoretical estimates (c). Similarly, the laser amplitude scaling for fixed pulse duration of 4 cycles and different beam waists of 3, 5, and 7 $\mu$m are also compared with analytical results (d). Here, for all the results we have used the energy of the electron bunch to be $\gamma_i = 30$. The points in (c) and (d) represent the result of full test particle simulations, and solid lines represent the scalings predicted by Eq. \ref{thnorr}.}
\label{fig1}
\end{figure}
Similarly, the longitudinal component of the force can be written as: $dp_z/dt = - \upsilon_x \pddz{A_x}$, however the rate change of energy is given by $d\gamma/dt = -\upsilon_x \pddt{A_x}$. In the coordinate system moving with the laser pulse, it can be easily shown that, $\zeta_z \equiv \gamma - p_z$ would be a constant of the motion, whose value can be obtained by imposing the  condition that initially $p^i_{z} = - \gamma_i \upsilon^i_{z} \sim - \gamma_i$, where $\gamma_i$ is the \textit{initial} energy of the electron bunch and minus indicates the counter-propagation of the electron bunch. This initial condition leads to $\zeta_z \sim \gamma_i + \gamma_f$, where $\gamma_f$ is the \textit{final} energy of the particle as the particle leaves the interaction region. Generally, $\gamma_f \sim \gamma_i$ when there is no loss of  energy through RR. However, when RR can not be ignored, the energy of the outgoing electron bunch would be smaller than the energy of the incident electron bunch, which will be discussed later in the manuscript.

Next, using the definition of $\eta = t - z + \phi_0$, one can write $d\eta/dt = 1 - p_z/\gamma$, which implies  $\gamma d\eta/dt = \zeta_z \sim \gamma_i + \gamma_f$. There will be no prominent transverse deflection of the electron until it is near the Rayleigh range of the focused laser. So effectively the $x_0$ can be considered as the initial position of the particle at $z = z_r$. With these approximations, Eq. \ref{ptildeeta} is simplified to be
\be \DDTa{\tilde{p}_x} \sim \frac{2}{(\gamma_i + \gamma_f)} \frac{A_x^2}{2 w_0^2} 
\Big[x_0 + \frac{A_x \eta}{\gamma}\Big] \label{ptildeeta2}\ee   
Terms with $\eta A_x^3$ vanish upon integration whence Eq. \ref{ptildeeta2} can be easily integrated over $\eta$ to obtain $\tilde{p}_x$
%\be \tilde{p}_x \sim \frac{1}{\gamma_f}\frac{x_0}{2 w_0^2} \int\limits_{-\infty}^{\infty} A_x^2\ d\eta  \ee
\be \tilde{p}_x \sim \frac{a_0^2}{(\gamma_i+\gamma_f)}\frac{x_0}{2 w_0^2}\exp\Big[\frac{-r_0^2}{w_0^2}\Big]
\int\limits_{-\infty}^{\infty} \exp\Big[\frac{-8 \ln(2) \eta^2}{\tau_0^2}\Big] \cos^2(\eta) d\eta  \ee
\be \tilde{p}_x \sim \sqrt{\frac{\pi}{32\ln(2)}} \frac{a_0^2\tau_0}{(\gamma_i + \gamma_f)}
\frac{x_0}{2 w_0^2}\exp\Big[\frac{-r_0^2}{w_0^2}\Big] \label{pxf0}.\ee
There are manifold scattering angles associated with each individual electron in the bunch, but the width of the electron bunch on the detector, viz. the most easily accessible signal, is set by the largest scattering angle. This largest scattering depends on the maximum transverse momentum. From, Eq. \ref{pxf0} it can be deduced that the $\tilde{p}_x$ would maximize for an electron with initial position $x_0 = w_0/\sqrt{2}$ by setting $d\tilde{p}_x/dx_0 = 0$, and hence the maximum net momentum along $x$ can be written as $\tilde{p}_{x;max} = \tilde{p}_x|_{x_0=w_0/\sqrt{2}}$ and is given by,
\be\tilde{p}_{x;max} \sim \sqrt{\frac{\pi}{256\ \ln(2)\ \text{e}}}\ \ \frac{a_0^2}{(\gamma_i +\gamma_f)}\frac{\tau_0}{w_0}.\ee 
The maximum scattering angle of the electron bunch (as estimated from the $z$ axis) would be; $\tan \Theta = \tilde{p}_{x;max}/p_z$. As discussed earlier, after the scattering $p_z \sim \gamma_f$ and hence the scattering angle is estimated to be
\be \tan\Theta \sim \kappa \sqrt{\frac{\pi}{256\ \ln(2)\ \text{e}}}\ \ \frac{a_0^2}{\gamma_f(\gamma_i+\gamma_f)}\frac{\tau_0}{w_0}.\label{thnorr}\ee 
Here we introduced the fitting parameter $\kappa$ to account for stronger scattering in a realistic laser field due to higher order field correlations, unaccounted for in our simple model.
The constant scaling factor for all the theoretical results presented in this manuscript is $\kappa \equiv 2\sqrt{2}$, which is derived from the benchmarking against numerical simulations. Now, Eq. \ref{thnorr} determines the maximum width of the pondermotively scattered electron bunch. The RR is neglected thus far, in the later part of the manuscript we will discuss the scenarios when the RR plays an crucial role, and how we can quantify the same.   
%, it might be because the actual temporal Gaussian envelope is approximately $2\sqrt{2} \tau_0$. 
 
\begin{figure}[b]
\centering\includegraphics[totalheight=0.8\columnwidth]{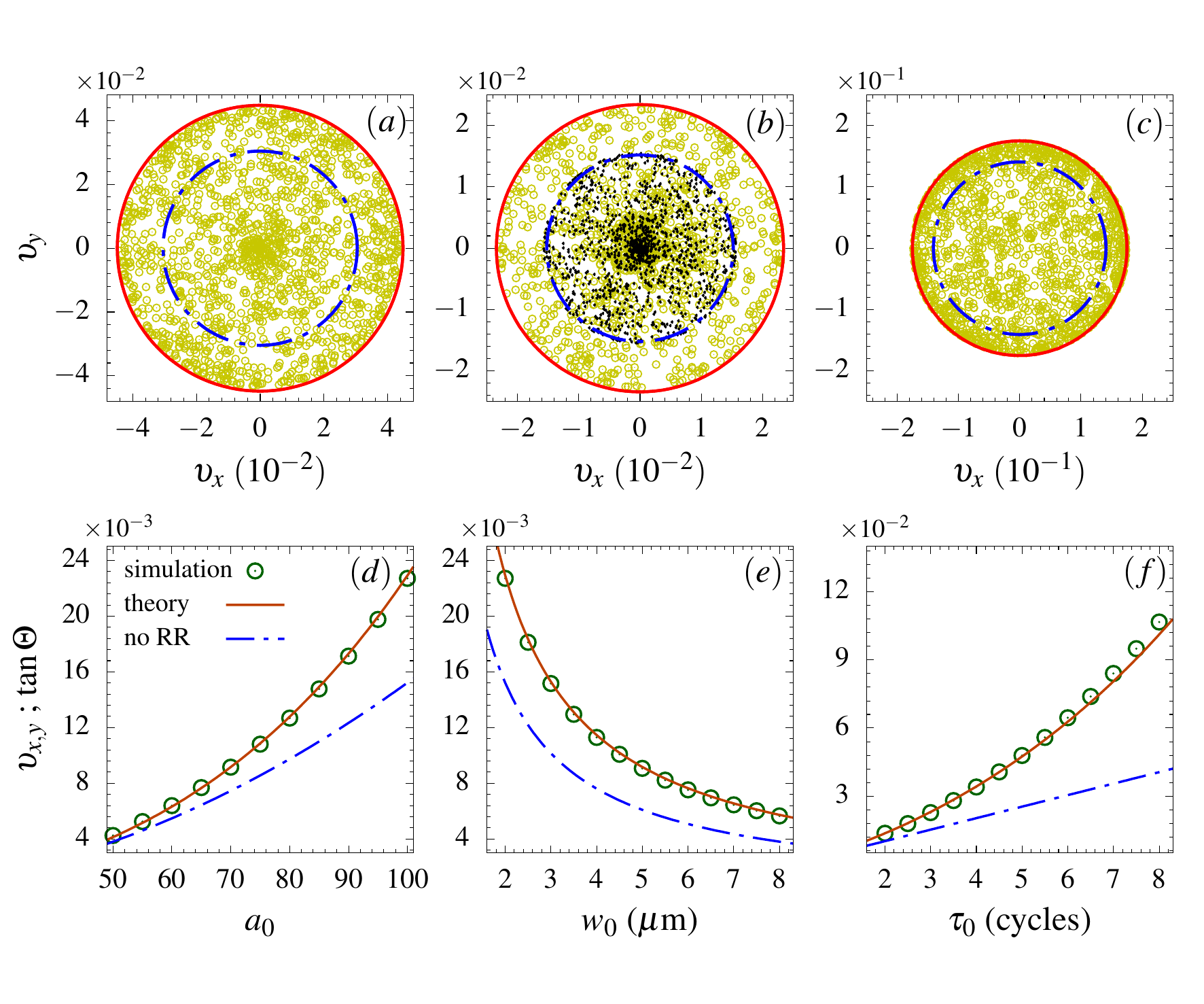}
\caption{The simulated final transverse velocity of outgoing electrons are compared with the predictions of the Eq. \ref{thnorr} for various interaction parameters [$a_0$,$\tau_0$ (cycles),$w_0$($\mu$m),$\gamma_i$]: (a) [100,4,3,200], (b) [100,3,2,300], and (c) [100,5,4,90]. The outer red circle represents the theoretically predicted scattering angle when the RR is included, however the inner dashed blue circle is the case when RR is ignored i.e. $\gamma_f = \gamma_i$ is considered in Eq. \ref{thnorr}. The black dots in (b) are for the case when the RR is forcefully disabled in the simulation. Next, we present and compare the scaling of the: (d) laser peak amplitude for a fixed $\tau_0 = 3$ cycles and $w_0 = 2 \mu$m ; (e) beam waist for fixed $a_0 = 100$ and $\tau_0 = 3$ cycles ; and (f) laser pulse duration for fixed $a_0 = 100$ and $w_0 = 2 \mu$m (refer text for $\gamma_f$ used in these scaling laws). We have used $\gamma_i = 300$ for the results presented in (d), (e) and (f). The points (small circles) in all the scaling curves denote the results from the simulation and solid line represent the estimation from the Eq. \ref{thnorr}, and dashed blue line is case when RR is disabled in Eq. \ref{thnorr} i.e. $\gamma_f = \gamma_i$ is considered.  }
%\caption{(a) $a_0 = 100$, $\tau_0 = 4$T, $w_0 = 3 \mu$m, $\gamma_i = 200$. (b) $a_0 = 100$, $\tau_0 = 3$T, $w_0 = 2 \mu$m, $\gamma_i = 300$. (c) $a_0 = 100$, $\tau_0 = 5$T, $w_0 = 4 \mu$m, $\gamma_i = 90$. (d) $\tau_0 = 3$T, $w_0 = 2 \mu$m, $\gamma_i = 300$. (e) $\tau_0 = 3$T, $a_0 = 100$, $\gamma_i = 300$. (f) $w_0 = 2 \mu$m, $a_0 = 100$, $\gamma_i = 300$.}
\label{fig2}
\end{figure}

\begin{figure}[b]
\includegraphics[trim=0 0 80 0,totalheight=0.95\columnwidth]{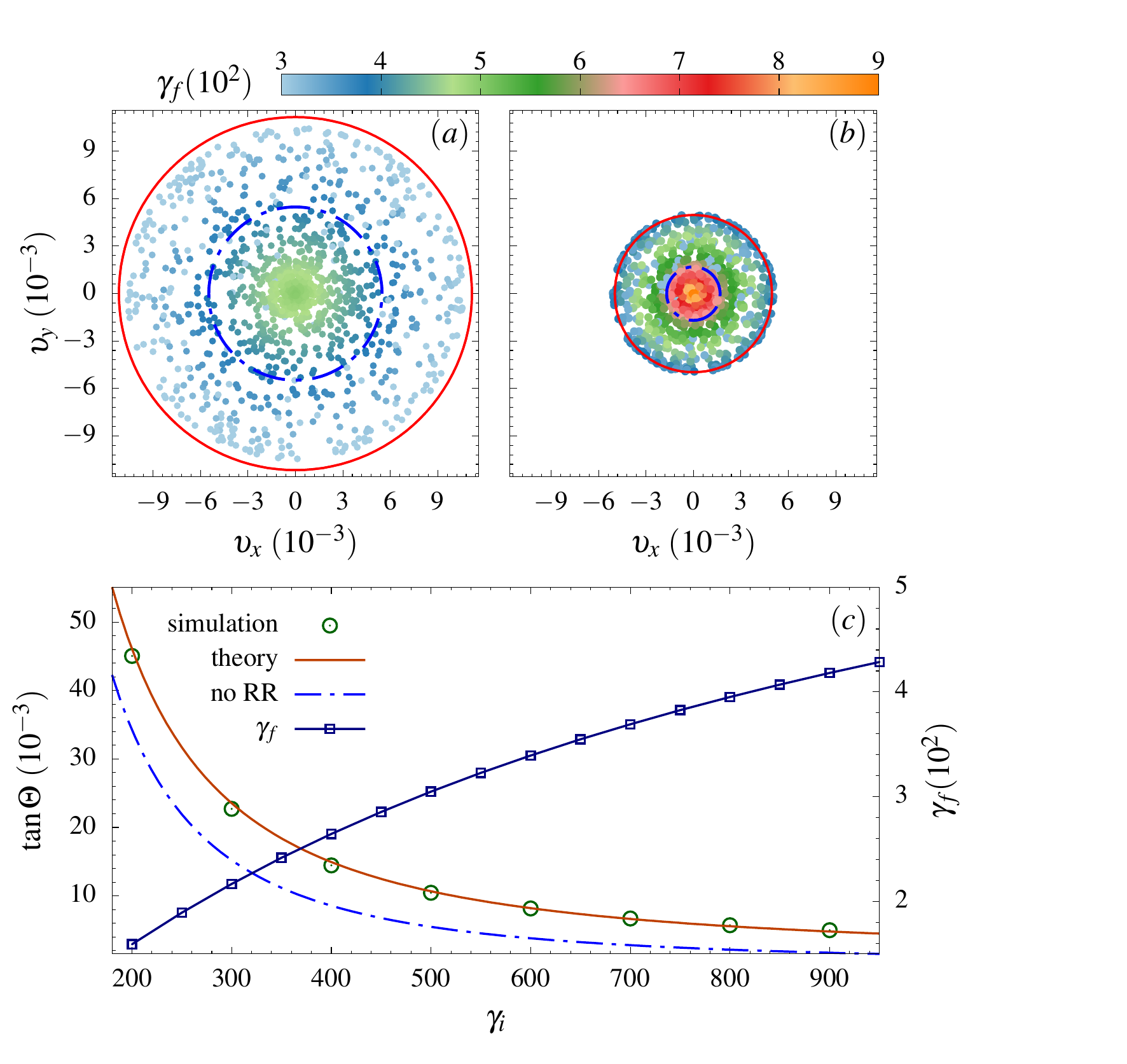}
\caption{The numerically calculated final transverse velocity of electrons for 3 cycle laser with peak amplitude $a_0 = 100$, and beam waist 2 $\mu$m are calculated by using $\gamma_i = 500$ (a) and $\gamma_i = 900$ (b). The color coded points in (a) and (b) represent each test particle with their respective $\gamma_f$, and the outer sold red (inner dashed blue) circle is theoretically estimated scattering angle with RR (without RR). The scaling of the scattering angle with the incoming electron bunch energies is presented in (c), for 3 cycle laser pulse with $a_0 = 100$, and $w_0 = 2 \mu$m. Open circles (solid orange line) represent the results from the full test particle simulations (theoretical calculation). The outgoing energy of the single electron ($\gamma_f$) as if it interacts with the plane wave of amplitude $a_0 \exp[-0.25]$ is shown by the line with points on right $y$ axis. These estimates of $\gamma_f$ are actually used in Eq. \ref{thnorr} for the scaling (solid orange line) in this figure. The dashed blue line is case when RR is ignored i.e. $\gamma_f = \gamma_i$ is used.   }
\label{fig3}
\end{figure}
 
In order to benchmark the predictive power of Eq. \ref{thnorr} to predict the scattering angle of the outgoing electron bunch, we carried out relativistic test particle simulations \cite{10.1063/1.4932995,Felix2019_NJP,10.1038/s41598-019-55949-3}. We modeled the laser field as a focused paraxial beam. The paraxial description is a perturbative expansion which satisfies the Maxwell equations to a given order in the diffraction limit $\theta_0 = w_0/z_r$ (we retain the expansion terms upto $\theta_0^5$) \cite{PhysRevSTAB.5.101301,PhysRevAccelBeams.19.094701}. The effect of radiation reaction is included in the dynamics through the perturbative approximations of the Landau Lifshitz (LL) force equation \cite{Tamburini:2010,10.1007/s41614-020-0042-0} and the electrons are pushed using the standard, numerically stable, Boris leapfrog algorithm \cite{Filippy-2001}. For all the simulations the electron bunch is considered to be a circular disk of radius (width) $2\lambda$ ($\lambda/2$), with $\lambda$ being the laser wavelength. The scattering angles of the electron bunch are not sensitive to the bunch dimensions, however for larger bunch radius one needs more number of test particles for any significant statistics on the scattering angles. For all the simulation results presented we have considered 1500 randomly distributed electrons in a circular disk. The radiation reaction is always included in all the simulations, unless specified.

In Fig. \ref{fig1} we have compared the scaling of the scattering angles with peak laser amplitude ($a_0 \leq 30$) for different pulse duration and the beam waist. For this comparison, we have used $\gamma_i = 30$. These parameters are chosen such that the RR is not crucial for the interaction dynamics ($\gamma_f \sim \gamma_i$) . The scaling laws predicted by Eq. \ref{thnorr} (solid lines) are in excellent agreement with realistic and relativistic test particle simulations (points on curves). The details of all the parameters are mentioned in the caption of Fig. \ref{fig1}. Furthermore in Fig. \ref{fig1}(a) and (b), we have also presented the final bunch transverse velocities for laser amplitudes $a_0 = 3$ and $a_0 = 30$ with FWHM pulse duration of 5 cycles and beam waist of 2 $\mu$m. The dots represent each test electron, and solid circle represents the theoretical estimates of Eq. \ref{thnorr}. Here, it should be noted that there is no strict condition on the ratio $a_0/\gamma_i$ for the valid characterization. However, if the incoming electron energy is very small as compared to the peak laser amplitude i.e. $a_0/\gamma_i \gg 1$, then the counter-propagating electrons tend to be reflected along the laser propagation direction, and the theoretical estimates of Eq. \ref{thnorr} will not be valid. Overall, the condition $\tan\Theta < 1$ (not necessary equivalenet to $a_0 < \gamma_i$) ensures an accurate measurement of the scattering angles.       

For ultra-intense laser beams RR implies radiative energy loss of the electrons. In the context of the pondermotive scattering of the electron bunch, the detailed account of the interaction dynamics is not crucial, but the knowledge of final energy of the electron ($\gamma_f$) after the exit from the laser focus suffices to estimate the extent of the scattering. The dominance of the RR in the interaction dynamics is generally measured in terms of the parameter $R \sim 4 r_0 a_0^2 \gamma_i/3$, where $r_0 = k r_e \sim 2.21\times 10^{-8}$ (for 800 nm laser)  is the dimensionless classical electron radius with $r_e \equiv e^2/(4\pi\epsilon_0 m_e c^2)$ (in SI units) \cite{10.1007/s41614-020-0042-0}. The parameter $R$ is calculated by comparing the energy loss per cycle as compare to the electron rest mass energy \cite{Harvey:2011dp,10.1063/1.4932995,DiPiazza:2009RR}. For example, $R \sim 8\times 10^{-4}$ for the case of $a_0 = \gamma_i = 30$ in Fig. \ref{fig1}, and hence the estimates of Eq. \ref{thnorr} with $\gamma_f = \gamma_i$ are in agreement with the simulations, wherein RR is always enabled. 

It should be noted that the Eq. \ref{thnorr} predicts the scattering angle of electrons pondermotively scattered electrons by ultra-intense laser pulses. The scattering angle depends on the peak laser amplitude, laser pulse duration and beam waist, along with the incoming and outgoing electron energy. The overall form of Eq. \ref{thnorr} is quite generic, as can be inferred from Fig. \ref{fig1} wherein the energy of the outgoing electron is the same as that of the incoming electron because of negligible RR. However, when RR is dominant, the electron(s) would lose their energy in the form of radiation. It should be also noted that this emitted radiation (Thomson/Compton scattering) can also be used as a diagnostic tool to characterize the laser pulse \cite{10.1103/physrevaccelbeams.21.114001, 10.1038/s41598-019-55949-3}. However,  in our setup, the energy lost by the electrons due to RR would clearly manifest in the scattering angles, as a lower $\gamma_f$ implies higher scattering angle [cf. \ref{thnorr}]. In order to estimate the scattering angle of the pondermotively scattered electron, the detailed dynamics of the electron in the laser focus is not relevant, rather the final exit energy as it leaves the laser focus would be sufficient for the prediction of the scattering angle.

The electron bunch dynamics in focused laser fields with the inclusion of the radiation-reaction is too involved to be solved through analytical formulation. However, the dynamics of an electron in a plane wave with RR has been extensively studied in the past \cite{10.1007/s11005-008-0228-9, 10.1103/physrevd.84.116005, 10.1103/physreve.88.011201, 10.1007/s41614-020-0042-0}. We already discussed that the electron at a transverse distance $r_0 = w_0/\sqrt{2}$ would be scattered most, and so in order to estimate the scattering angle of the electron bunch with RR, it would be sufficient to approximate the final energy of that electron. We can thus obtain the $\gamma_f$ of the electron by estimating the energy loss due to RR of a single particle at $r_0 = w_0/\sqrt{2}$ interacting with a \textit{plane wave} having the maximum field strength at the Rayleigh range. The maximum field strength accessible to the charge particle at $r_0 = w_0/\sqrt{2}$ and $z = z_r$ is $\sim a_0 \exp[-0.25]$. In a real focused laser field, however, the electron would scatter away, thus experiencing lower RR and hence the final exit energy would be slightly higher than estimated via this simple plane wave model. We will see that this does not affect the predictive power of our model.

In Fig. \ref{fig2}, we compare the transverse velocity and scattering angle of outgoing electrons with the predictions of the Eq. \ref{thnorr} for various interaction parameters [refer caption of Fig. \ref{fig2}]. The final transverse velocities presented in Fig. \ref{fig2} (a), (b) and (c) show the theoretically predicted scattering angle with RR (outer red circle) and without RR (inner blue circle). In all these cases the $\gamma_f$ is calculated by numerically solving the electron dynamics in a plane wave of intensity $\sim a_0 \exp[-0.25]$, which eventually is just an input to the Eq. \ref{thnorr}. In Fig. \ref{fig2}(b) we also carried out a simulation with forcefully disabled RR, and the transverse velocity of individual electron (black dots) are very well captured by  Eq. \ref{thnorr} (blue circle) with $\gamma_i = \gamma_f$. It should be noted that in Fig. \ref{fig2}(c), the interaction of the electron bunch with $\gamma_i = 90$ with a laser of peak amplitude $a_0 = 100$ is presented, and the estimated scattering angle even in this case ($\gamma_i < a_0$) is well reproduced. As we mentioned earlier, there is no strict condition on the ratio of $a_0/\gamma_i$, but the overall $\tan\Theta$ should be small or the electrons tend to reflect toward the laser propagation direction. 

The scaling for the laser amplitude is compared in Fig. \ref{fig2}(d), wherein the peak laser amplitude is varied from the $a_0 = 50$ to $a_0 = 100$. The theoretical value of the scattering angle is estimated by Eq. \ref{thnorr}. The final exit energy $\gamma_f$ is estimated by solving the dynamics of a single electron at $r_0 = w_0/\sqrt{2}$ and $z=z_r$ within the LL model in a plane wave of accordingly reduced intensity for this parameter. The result is found as $\gamma_f(a_0) = A \sqrt{R(a_0,\gamma_i) \tau_0} + B$, where $R(a_0,\gamma_i) \sim 4 r_0 a_0^2 \gamma_i/3$ is related to the energy loss of the electron in a laser cycle. The scaling law for the beam waist for a given peak laser intensity and pulse duration is presented in Fig. \ref{fig2}(e). As we are estimating the final exit energy by studying the plane wave interaction at the reduced intensity at Rayleigh range and $r_0 = w_0/\sqrt{2}$ the final energy $\gamma_f$ would be unaffected by variations of the beam waist. This facet enabled us to use the same $\gamma_f(a_0)$ as we used for Fig. \ref{fig2}(d) [for other details please refer the caption of Fig. \ref{fig2}]. 

Next, the scaling with laser pulse duration for fixed peak amplitude and beam waist is presented in Fig. \ref{fig2}(e). The determination of the laser pulse duration plays a crucial role in various applications \cite{10.1038/s41598-019-55949-3}. In order to compare the scattering angle with laser pulse duration, in Fig. \ref{fig2}(e) we have again estimated the final exit energy of the electron by numerically solving the \textit{single} electron dynamics in a plane wave of reduced intensity as discussed earlier. The FWHM dependent final exit energy is then fitted through, $\gamma_f(\tau_0) = C \sqrt{R(a_0,\gamma_i) \tau_0} + D$. In Fig. \ref{fig2}(c), (d) and (e) we have also presented the scaling laws when the RR is disabled i.e. $\gamma_f = \gamma_i$ is used in Eq. \ref{thnorr}. It can be observed from the Fig. \ref{fig2}(e), that the energy loss through RR manifests in the deviation from a linear variation of scattering angle with the laser pulse duration [refer Eq. \ref{thnorr}]. The deviation from the linear scaling arises from the estimation of the final exit energy of the electron through: $\gamma_f(\tau_0) \propto \sqrt{\tau_0}$ [for other details please refer the caption of Fig. \ref{fig2}]. A detailed account of the energy loss through RR as function of the laser pulse duration is beyond the scope of the present work. However, the theoretical understanding of the radiation dominated interaction dynamics of the electron in the plane wave may be crucial in this aspect \cite{10.1007/s11005-008-0228-9, 10.1103/physrevd.84.116005, 10.1103/physreve.88.011201, 10.1007/s41614-020-0042-0}. 

So far we have discussed how the pondermotively scattered electrons can be utilized as a potential diagnostic tool for ultra-intense laser pulses, where the traditional optical setup will not be suitable. On the contrary, the proposed setup can also be explored for studying the different models for the radiation reaction. The energy loss through a particular RR model can be verified experimentally by measuring the scattering angles of the given electron bunch after it undergoes pondermotive scattering. In Fig. \ref{fig3}, we present the final transverse velocity distribution of the electron bunch for the laser with peak amplitude $a_0 = 100$, pulse duration of 3 cycles and beam waist of 2 $\mu$m. The final transverse velocities of the electrons are presented for two cases when the incoming electron energy is $\gamma_i = 500$ (a) and $900$ (b) respectively. The final exit energy of a single electron at $r_0 = w_0/\sqrt{2}$ at $z=z_r$ is calculated by considering the reduced intensity in the plane wave ansatz. The case when  RR is not considered is also shown by an inner blue dashed curve. Even in this RR dominated regime, the predictions of Eq. \ref{thnorr} along with the plane wave ansatz are found to be accurate. The scaling of the scattering angle with the incoming electron bunch energies is also compared in the Fig. \ref{fig3}(c), and the agreement of theoretical estimation of Eq. \ref{thnorr} with full test particle simulation is found to be encouraging. The final exit energy of the electron bunch is also color coded in the Fig. \ref{fig3}(a) and (b) along with the variation of $\gamma_f$ with $\gamma_i$ is shown in Fig. \ref{fig3}(c). 

In conclusion, we propose an analytical formulation for the scattering angles of ponderomotivly scattered relativistic electrons. The proposed expression is able to quantitatively predict the scattering angle depending on the peak laser amplitude, it's FWHM pulse duration and also it's beam waist. The generic nature of the proposed expression is also found to be valid when  radiation reaction is prevalent. Not only will this setup and formulation be useful for the  characterization of  ultra-intense laser pulses, but also the theoretical studies of radiation reaction may benefit by its experimental realization.

%The characterization of the ultra-intense laser pulses is always challenging, and most of the time the direct characterization seems to be not feasible because of the damage caused to optical elements. An experiment is generally setup to probe only a single physical property of the laser pulse, be it a peak laser amplitude, it's FWHM pulse duration or the beam waist.  In this letter we propose an analytical formulation for the scattering angles of ponderomotivly scattered relativistic electron bunch. The proposed expression is able to quantitatively predict the scattering angle depending on the peak laser amplitude, it's FWHM pulse duration and also it's beam waist. The theoretical estimates for the scattering angles and associated scaling laws are benchmarked with full relativistic test particle simulations in realistic focused laser fields. The generic nature of the proposed expression is also found to be valid when the radiation reaction is prevalent. In radiation dominated regime, we need to estimate the final exit energy of the electron, which will be an input to the proposed formula for scattering angle. The final energy of the electron after the RR is modeled through a plane wave ansatz in reduced peak laser intensity. Not only this setup and formulation will be useful for characterization of the ultra-intense laser pulses, but the theoretical studies in the field of the radiation reaction may also be benefited by it's experimental realization.  

\textit{Acknowledgments.}- A.H. would like to acknowledge the computing resources funded by the DST-SERB, Government of India, through the project EMR/2016/002675. 

\bibliographystyle{apsrev4-1}
%\bibliography{Bibliography}

 
\end{document}